**Peer Review** The peer review history for this article is available as a PDF in the Supporting Information.

**Key Points:**
- The midlatitude jet-streams and meridional circulation cells of the Jovian atmosphere are reproduced in a 3D, high-resolution, deep numerical model
- The eddy momentum fluxes drive the jets and circulation cells in the midlatitudes, consistent with Ferrel cell theory adjusted to gas giants
- A stacked meridional circulation pattern emerges under certain conditions, consistent with Juno microwave radiometer measurements of the Jovian atmosphere

**Supporting Information:**
Supporting Information may be found in the online version of this article.

**Correspondence to:**
K. Duer,
keren.duer@weizmann.ac.il

**Author Contributions:**
**Conceptualization:** Keren Duer, Eli Galanti, Yohai Kaspi
**Formal analysis:** Keren Duer
**Investigation:** Keren Duer
**Methodology:** Keren Duer, Yohai Kaspi
**Software:** Keren Duer
**Supervision:** Yohai Kaspi
**Visualization:** Keren Duer
**Writing – original draft:** Keren Duer

# Gas Giant Simulations of Eddy-Driven Jets Accompanied by Deep Meridional Circulation

Keren Duer[1] , Eli Galanti[1] , and Yohai Kaspi[1]

[1]Department of Earth and Planetary Sciences, Weizmann Institute of Science, Rehovot, Israel

**Abstract** Jupiter's atmosphere comprises several dynamical regimes: the equatorial eastward flows and surrounding retrograde jets; the midlatitudes, with the eddy-driven, alternating jet-streams and meridional circulation cells; and the jet-free turbulent polar region. Despite intensive research conducted on each of these dynamical regimes over the past decades, they remain only partially understood. Saturn's atmosphere also encompasses similar distinguishable regimes, but observational evidence for midlatitude deep meridional cells is lacking. Models offer a variety of explanations for each of these regions, but only a few are capable of simulating more than one of the regimes at once. This study presents new numerical simulations using a 3D deep anelastic model that can reproduce the equatorial flows as well as the midlatitudinal pattern of the mostly barotropic, alternating eddy-driven jets and the meridional circulation cells accompanying them. These simulations are consistent with recent Juno mission gravity and microwave data. We find that the vertical eddy momentum fluxes are as important as the meridional eddy momentum fluxes, which drive the midlatitudinal circulation on Earth. In addition, we discuss the parameters controlling the number of midlatitudinal jets/cells, their extent, strength, and location. We identify the strong relationship between meridional circulation and the zonal jets in a deep convection setup, and analyze the mechanism responsible for their generation and maintenance. The analysis presented here provides another step in the ongoing pursuit of understanding the deep atmospheres of gas giants.

**Plain Language Summary** Jupiter's atmosphere has different jet-stream patterns in different regions. However, they are still not fully understood, even though they have been a subject of interest for a long time. Saturn's atmosphere is similar, but information is lacking about the 3D wind structure. In this study, we use a deep general circulation model that simulates the atmosphere of a gas giant. The simulations reproduce both the equatorial winds and the jet-streams in the midlatitudes, and conceptually match recent measurements from the Juno and Cassini spacecraft. The model shows that turbulence drives the 3D structure of the jet streams. We discuss the parameters controlling the number of jets and cells, their extent, strength, and location. We find a strong connection between the circulation moving north-south and up-down and the east-west jet streams, and we examine how they are created and maintained. This study reveals how the jet-streams on gas giants like Jupiter and Saturn behave below the visible cloud layer.

## 1. Introduction

The atmospheres of Jupiter and Saturn have been observed and studied for centuries. Their turbulent nature is clear even in visible light, but it is still uncertain what role turbulence takes in the deep dynamics of the planets. The imaging instrument onboard the Cassini spacecraft has provided a clear linkage between the atmospheric mean flow and the cloud-level disturbances (eddies, deviations from the mean flow) in both planets (Del Genio & Barbara, 2012; Salyk et al., 2006), hinting at the potential similarities between the atmospheres of gas giants and those of the terrestrial planets. In the midlatitudes of Earth, for example, the momentum balance is dominated by the eddies, which maintain the eddy-driven jets and the meridional circulation cells (Chemke & Kaspi, 2015; Lachmy & Kaspi, 2020; Schneider, 2006; Vallis, 2017). The cloud-level measurements of Jupiter and Saturn suggest a similar relation, as the midlatitudes of both planets reveal a strong correlation between the zonal wind and the eddy fluxes (Figures 1a and 1b).

The gravity experiments conducted by the Juno spacecraft for Jupiter (Iess et al., 2018; Kaspi et al., 2018) and the Cassini Grand Finale for Saturn (Galanti et al., 2019; Iess et al., 2019) have resulted in estimates for the penetration depth of the atmospheric mean flow (Figures 1c and 1d, blue). The common value for the depth of the Jovian jets is about 3,000 km, and the Saturnian jets is about 9,000 km (Kaspi et al., 2020). These depths can

**Writing – review & editing:** Eli Galanti, Yohai Kaspi

be projected onto the 1 bar pressure level to determine the latitude of the tangent cylinder, which is an imaginary line that touches the planet's interior at the equatorial plane and runs parallel to its rotation axis. A combination of gravity and magnetic measurements yield similar, but more barotropic wind decay profiles (Figures 1c and 1d, red, Galanti and Kaspi (2021)). To estimate the jets' deep structure, a jet aspect ratio can be defined as $A = \frac{WR}{D}$, where $W$ is the averaged midlatitudinal jet latitudinal width (in radians), $R$ is the planet's radius and $D$ is the atmospheric depth. Hence, the atmosphere-to-radius ratio is 5% (15%), and the average jet aspect ratio is 2 (0.8) on Jupiter (Saturn). For comparison, on Earth, assuming that the weather layer is roughly 10 km deep, the atmosphere-to-radius ratio is 0.15% and the jet aspect ratio is 223. Considering these massive atmospheres, all shallow approximations, which are usually made when analyzing the atmospheres of the terrestrial planets, cannot hold for gas giants. Moreover, as the jets decay over thousands of kilometers, they can be considered as relatively barotropic, unlike the baroclinic terrestrial jets. Furthermore, the absence of a solid surface and the distinct major energy sources are additional differences that must be considered for understanding the dynamics of gas giants.

While the Jovian jets were shown to be eddy-driven since the Cassini flyby (Salyk et al., 2006), the mean meridional circulation remained uncertain until recently due to a lack of measurements beneath the cloud level (Fletcher et al., 2020). Measurements by the microwave radiometer (MWR) onboard the Juno spacecraft (Bolton et al., 2017; Li et al., 2017; Oyafuso et al., 2020) have verified the existence of prominent, deep penetrating, meridional circulation cells in Jupiter's midlatitudes (Duer et al., 2021). These cells resemble the Ferrel cells on Earth, as they are driven by a convergence of meridional eddy momentum fluxes and are accompanied by a strong barotropic jet stream. The Ferrel cell is essentially the average movement of air in the midlatitudes of Earth. It exists to counterbalance the convergence of the eddies in that region (Vallis, 2017). While an accurate quantification of the Jovian vertical velocities is unavailable due to nonexistent measurements, Duer et al. (2021) estimated their relative strength and position according to the MWR and eddy momentum flux measurements (Figure S1 in Supporting Information S1). While observational evidence exists regarding the Ferrel-like cells, some of their properties are still uncertain. How deep are these cells? The Juno MWR measurements provided evidence only for the upper 300 km of the atmosphere, but the cells might penetrate much deeper. What closes the cells at the bottom instead of the surface drag, which is an important component in the circulation on Earth? Theoretical and numerical studies imply that the Lorentz force and/or a stable layer could potentially dissipate the jets in the interior region (e.g., Gastine & Wicht, 2021; Liu et al., 2008). What is the role of the eddies below the cloud level? And what are the leading order momentum and energy balances in gas giants' midlatitudes that allow multiple jets and circulation cells? In this study, we will give an insight into some of these questions using a numerical model that can reproduce the midlatitudinal pattern of the mostly barotropic, alternating, eddy-driven jets and the meridional circulation cells accompanying them.

Several numerical simulations have shown the structure of alternating jet streams and meridional cells at midlatitudes by both shallow and deep models. Early work regarding the Jovian jet streams generation mechanism includes forced 2D turbulence simulations (e.g., Williams, 1978), successfully demonstrating an inverse cascade in Jovian conditions, and unforced shallow-water simulations with small Rossby number, resulting in multiple jets from geostrophic turbulence (e.g., Cho & Polvani, 1996a, 1996b). Newer 3D simulations of a shallow atmosphere with parameterization of drag produce eddy-driven jets and circulation cells (e.g., Lian & Showman, 2008; Spiga et al., 2020). However, the jets are usually baroclinic and vanish at a depth of a few bar (Liu & Schneider, 2010; Schneider & Liu, 2009), which is inconsistent with recent estimations of the jets, showing that they are relatively barotropic and deep (Duer et al., 2020; Galanti & Kaspi, 2021; Galanti et al., 2019, 2021; Kaspi et al., 2018). Weather layer simulations with passive tracers also reveal multiple jets and circulation driven by eddy momentum fluxes, where the leading order momentum balance is similar to Ferrel cells balance. However, these simulations are shallow (∼20 bar) and their applicability to the massive atmospheres of gas giants may not be straightforward (Young et al., 2019a, 2019b).

Deep, convection-driven, 3D simulations can produce alternating jets, but the meridional circulation and the eddy momentum fluxes are rarely addressed and analyzed in such setups (e.g., Aurnou & Olson, 2001; Christensen, 2001; Christensen, 2002; Heimpel et al., 2005, 2016, 2022). In these types of simulations, boundary conditions (e.g., Aurnou & Heimpel, 2004; Jones et al., 2003; Yadav et al., 2013, 2016), numerical viscosity (e.g., Wicht et al., 2002), forcing scheme (e.g., Showman et al., 2011) and compressibility (e.g., Jones & Kuzanyan, 2009; Kaspi et al., 2009) were shown to be important for the appearance and stability of the jets. Simpler models are able to generate the jet-circulation cell patterns, but they are forced in a form that dictates the number and extent of the jets independently from other model parameters (Christensen et al., 2020), which cannot indicate the jet generating mechanism. Recent dynamo models suggest that the presence of a stably stratified layer promotes the emergence of multiple zonal jets at midlatitudes (e.g., Gastine & Wicht, 2021; Wulff et al., 2022; Yadav et al., 2022).

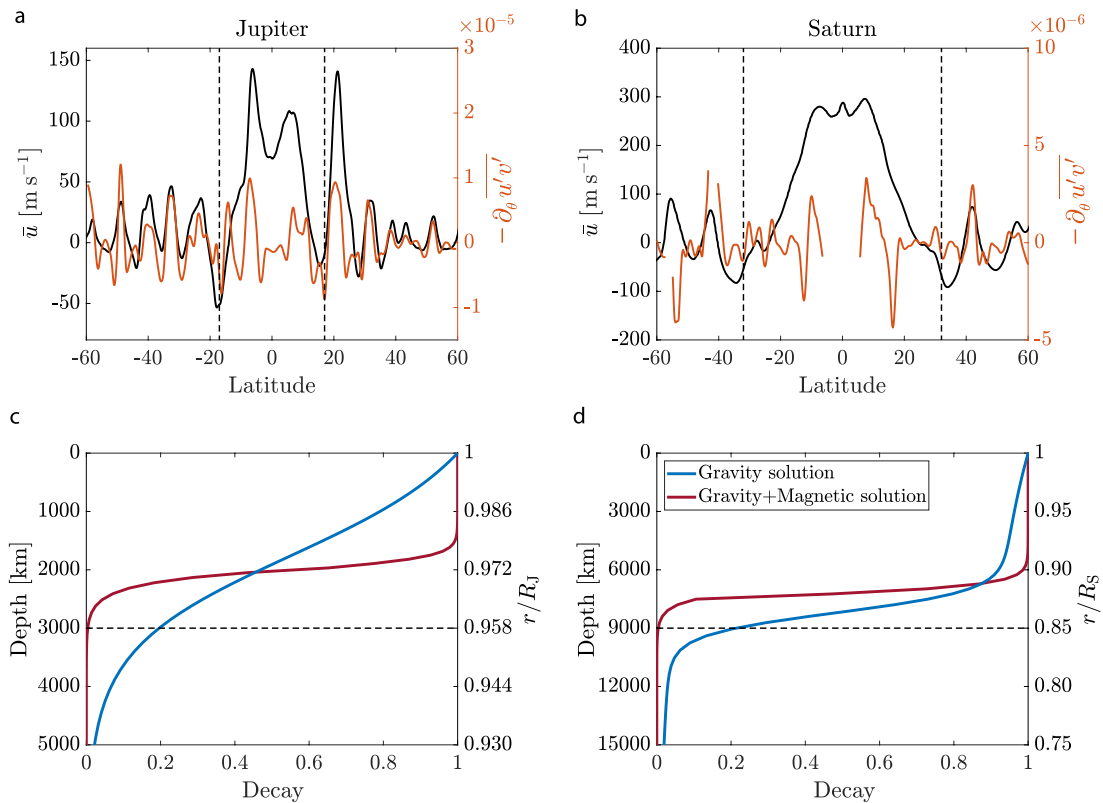

**Figure 1.** (a) Jupiter's and (b) Saturn's zonally averaged zonal cloud-level winds ($\bar{u}$ [m s$^{-1}$], black, Tollefson et al. (2017) and Garcia-Melendo et al. (2011), respectively) and the eddy momentum flux convergence at the cloud level recalculated from the Cassini imaging data ($\partial_\theta \overline{u'v'} \equiv \frac{1}{R\cos^2\theta} \frac{\partial(\cos^2\theta \overline{u'v'})}{\partial \theta}$ [m s$^{-2}$], orange, Salyk et al. (2006) and Del Genio and Barbara (2012), respectively). Saturn's wind field is displayed based on recent estimations of the rotation rate (Mankovich et al., 2019). The velocity lines are presented with a running average of 2° latitude and the eddy momentum flux lines are presented with a running average of 4° latitude such that small scale noise is not visible. Latitude is planetocentric. (c) and (d), The decay profile of the cloud-level zonal winds as a function of depth from the visible clouds (left ordinate) and normalized radius (right ordinate), that best fits the gravity measurements (blue, Kaspi et al. (2018) for Jupiter, Galanti et al. (2019) for Saturn), and the combined gravity-magnetic measurements (red, Galanti and Kaspi (2021)) from Juno and Cassini Grand Finale, respectively. The dashed black lines indicate the depth at which the wind speed reaches either 20% of the speed at the cloud level for the gravity solution or drops below 1% for the gravity-magnetic solution. These depths are situated at 3,000 km for Jupiter (c) and 9,000 km for Saturn (d). These depths are projected onto the surface to give the latitude of the tangent cylinder: 17° for Jupiter (dashed lines, panel a) and 32° for Saturn (dashed lines, panel b).

In this study, we use an anelastic, turbulent, fast-rotating and deep setup to capture the jet related features of gas giants. We provide an overview of the numerical equations, the leading order balances and the model setup in Section 2, followed by a presentation and analysis of the results in Section 3, with a detailed descriptions of the eddy-driven jets in Section 3.1, the zonal momentum equation leading balances in Section 3.2, and the meridional circulation cells in Section 3.3. We conclude in Section 4.

## 2. Numerical Simulations

### 2.1. Formulation

For simulations of gas giants atmospheres we use the open-source Rayleigh convection model, based on the Navier-Stokes equations (Featherstone et al., 2022). This is a commonly used model for dynamo general circulation simulations. The set of equations follows the general formulation used in the hydrodynamics and magnetohydrodynamics benchmarks presented by Jones et al. (2011).

The momentum equation can be written as

$$\frac{\partial \mathbf{u}}{\partial t} = \mathbf{u} \times \boldsymbol{\omega} - 2\boldsymbol{\Omega} \times \mathbf{u} - \nabla\left(\frac{p'}{\bar{\rho}} + \frac{1}{2}\mathbf{u}^2\right) + \mathbf{F}_\nu - \mathbf{g}\frac{S}{c_p}, \quad (1)$$

where $\mathbf{u} = (w\hat{r}, v\hat{\theta}, u\hat{\phi})$ is the velocity vector in spherical coordinates $(\hat{r}, \hat{\theta}, \hat{\phi})$, $t$ is time, $\boldsymbol{\omega} = \nabla \times \mathbf{u}$ is the vorticity, $\boldsymbol{\Omega}$ is the planetary rotation rate, $p$ is pressure, $\bar{\rho}(r)$ is the background-state density, $\mathbf{g}(r)$ is the gravitational acceleration, $\mathbf{F}_\nu$ is the viscous force, $S$ is the entropy and $c_p$ is the specific heat capacity at constant pressure. Bar represents a zonal average, and prime denotes deviations from this average, for example, $u = \bar{u} + u'$. Decomposing the vorticity and applying $\mathbf{F}_\nu = \nabla \cdot \mathbf{D}$ give the momentum equation as solved by many numerical codes:

$$\bar{\rho}\left(\frac{\partial \mathbf{u}}{\partial t} + \mathbf{u} \cdot \nabla \mathbf{u} + 2\boldsymbol{\Omega} \times \mathbf{u}\right) = \frac{\bar{\rho}}{C_p}\mathbf{g}S - \bar{\rho}\nabla\left(\frac{p'}{\bar{\rho}}\right) + \nabla \cdot \mathbf{D}, \quad (2)$$

where $\mathbf{D} = 2\bar{\rho}\nu\left(e_{ij} - \frac{1}{3}\nabla \cdot \mathbf{u}\right)$, $\nu$ is the constant kinematic viscosity, and $e_{ij}$ is the standard rate-of-strain tensor. Note that the Lorentz force is excluded as we do not consider magnetic field effects on the flow field. The continuity equation in an anelastic form is

$$\nabla \cdot (\bar{\rho}\mathbf{u}) = 0. \quad (3)$$

The equation for entropy can be written as

$$\bar{\rho}\bar{T}\left(\frac{\partial S}{\partial t} + \mathbf{u} \cdot \nabla S\right) = \nabla \cdot \kappa\bar{\rho}\bar{T}\nabla S + \Pi + Q_i, \quad (4)$$

where $\bar{T}$ is the background-state temperature, the energy flux is $-\kappa\bar{\rho}\bar{T}\nabla S$ (Braginsky & Roberts, 1995), $\kappa$ is the turbulent thermal diffusivity constant, $\Pi = 2\bar{\rho}\nu\left(e_{ij}e_{ij} - \frac{1}{3}(\nabla \cdot \mathbf{u})^2\right)$ is viscous heating, and $Q_i$ is the radially dependent internal heating (radiative heating or heating due to nuclear fusion) (Jones et al., 2011; Landau & Lifshitz, 1959). The three equations above (Equations 2–4) along with an equation of state, in this case a polytropic reference state (see Supporting Information S1), are a closed set of equations which can be solved numerically. We evolve the set of Equations 2–4 with a chosen background state using the Rayleigh code (Featherstone et al., 2022; Matsui et al., 2016). For further details regarding the Rayleigh model, the reference state, and a dimensionless notation please see Supporting Information S1.

## 2.2. The Leading Order Momentum Balance

In the midlatitudes of Earth, the eddies are responsible for the majority of the momentum and heat transport and the leading equations can be simplified to the Ferrel cell formulation (Vallis, 2017). In the gas giants, the leading order balance at depth has yet to be determined. Using similar approximations to the classical formulation of the Ferrel cell dynamics (zonal-average, steady-state), while considering deep atmospheres, the leading order momentum equation (Equation 2) becomes:

$$\bar{\rho}\left(\bar{\mathbf{u}} \cdot \nabla\bar{\mathbf{u}} + \overline{\mathbf{u}' \cdot \nabla\mathbf{u}'} + 2\boldsymbol{\Omega} \times \bar{\mathbf{u}}\right) = \frac{\bar{\rho}}{C_p}\mathbf{g}\bar{S} - \overline{\bar{\rho}\nabla_{\theta r}\left(\frac{p'}{\bar{\rho}}\right)} + \overline{\nabla \cdot \mathbf{D}}, \quad (5)$$

where $\nabla_{\theta r}$ is the gradient operator in the meridional and radial directions. Neglecting the mean advection terms (assuming the leading order is geostrophic) and considering that the Rossby number is small in the midlatitudes, the zonal momentum equation may be written (in spherical coordinates) as

$$\bar{\rho}\left(\overline{v' \cdot \partial_\theta u'} + \overline{w' \cdot \partial_r u'} - 2\Omega \sin\theta \bar{v} + 2\Omega \cos\theta \bar{w}\right) = \overline{\nabla \cdot \mathbf{D}}. \quad (6)$$

Finally, applying the continuity equation gives

$$\frac{1}{r\cos^2\theta}\frac{\partial}{\partial\theta}\left(\bar{\rho}\overline{u'v'}\cos^2\theta\right) + \frac{1}{r}\frac{\partial}{\partial r}\left(\bar{\rho}\overline{u'w'}r\cos\theta\right) - 2\bar{\rho}\Omega\sin\theta\bar{v} + 2\bar{\rho}\Omega\cos\theta\bar{w} = \overline{\nabla \cdot \mathbf{D}}. \quad (7)$$

Equation 7 is a complex three-term balance, between the full eddy momentum flux convergence, the full Coriolis force and the viscosity term. This balance is the ageostrophic force balance, which is dominant in the midlatitudes. Although Equation 7 is the leading order zonal momentum balance, under steady-state conditions, it does not include any component that directly relates to the mean zonal jets. Nonetheless, the evolution of zonal jets (being time-dependent) results from the convergence of momentum fluxes. The momentum cycle is such that momentum is transferred from the eddies to the mean flow through upgradient fluxes, resulting in the formation of jets, and, in turn, the mean flow dissipates, closing the momentum budget.

Equation 7 is somewhat different than the equivalent one describing Earth's midlatitudes. On Earth, the radial (or vertical, in cartesian coordinates system) eddy term is neglected due to the shallowness of Earth's atmosphere. Also, surface drag needs to be considered in the bottom boundary Ekman layer (e.g., Vallis, 2017). This leads to a three-term balance, between the meridional eddy momentum flux convergence, the Coriolis force and a drag term (usually a simple linear scheme is used). The upper branch of the Ferrel cell, above the Ekman layer, is characterized by a balance between the eddy momentum flux convergence and the Coriolis force:

$$f\bar{v} = \frac{1}{\bar{\rho} r \cos^2\theta} \frac{\partial}{\partial \theta}\left(\bar{\rho}\overline{u'v'}\cos^2\theta\right), \quad (8)$$

where $f = 2\Omega\sin\theta$. The lower branch (in an Ekman layer of depth $Z$) is dominated by a balance between the Coriolis force and the surface drag:

$$-fV \approx -\tilde{r}\tilde{u}_{\text{surf}}, \quad (9)$$

where $V = \int_0^Z \bar{\rho}\bar{v}dz$, $z$ is the vertical direction (parallel to the spin axis), and $\tilde{r}$ is a drag constant (Vallis, 2017).

For the gas giants, the equation takes a more complicated form. As the atmosphere is deep, the vertical terms in Equation 7 might be as significant as the meridional terms, resulting in an additional eddy term and the full Coriolis force. In the numerical framework, we must consider the viscosity terms as well (Equation 7). It is possible to take into account the impact of magnetic fields (the Lorentz force), which could influence the velocity field and might even replace the bottom drag term in the boundary layer (Duer et al., 2019; Gastine & Wicht, 2021; Kaspi et al., 2020; Liu & Schneider, 2010; Liu et al., 2008; Moore et al., 2019). However, for the purpose of this study, which is to investigate the effect of potential drag layers on the dynamics, we are not factoring in the magnetic field or drag terms. Including these can affect the zonal jets and the meridional circulation within and above the conducting layer. Numerical models that considered magnetic fields did not reveal a straight-forward Ferrel cells circulation pattern (e.g., Gastine & Wicht, 2021; Wulff et al., 2022). Instead, we demonstrate that a basic combination of boundary conditions supports the theoretical arguments presented above and validates Equation 7.

### 2.3. Numerical Setup

For analyzing small-scale turbulence-related features in gas giant atmospheres, we use high-resolution deep simulations, which can reproduce qualitatively similar characteristics at the outer boundary to those appearing in the visible cloud layer of Jupiter and Saturn. The domain depth is studied to allow midlatitude dynamics and the depth that is presented for further analysis is shallow enough to contain midlatitude dynamics, but deep enough to study radial eddies and their impact on the midlatitude zonal jets. We study the boundary conditions of a Jupiter-like simulation while all other parameters are kept constant. The control parameters of the simulation can be described by writing the equations in a dimensionless form (See Supporting Information S1), such that the main three parameters are the modified Rayleigh number $Ra^* = \frac{g_o \Delta S}{c_p \Omega^2 L}$, the Ekman number $Ek = \frac{\nu}{\Omega L^2}$, and the fluid Prandtl number $\text{Pr} = \frac{\nu}{\kappa}$. All of the simulations are solved with the same control parameters ($Ek = 5 \times 10^{-5}$, $\text{Pr} = 2$ and $Ra^* = 0.0132$), and Jupiter's radius and rotation rate. The range of radius ratios presented is $[0.35 - 0.84]$. Deep numerical simulations conducted for gas giants, here and in general, are intentionally over-forced and incorporate numerical viscosity that exceeds the molecular viscosity of Jupiter and Saturn by several orders of magnitude (Showman et al., 2011). This intentional over-forcing is necessary to counteract the high numerical viscosities required for such simulation setups while still maintaining high Rayleigh numbers. The simulations are calculated until dynamical steady-state is reached, and all the results below are shown for time-averaging of 300 rotations (except snapshots). Further details on the parameterization of the simulation and the background state are given in Supporting Information S1.

### 3. Results

#### 3.1. Eddy-Driven Midlatitude Jets

In simulations, the midlatitude jets (adjacent to the subequatorial retrograde jets) only appear when the depth of the simulation is limited enough to prevent the equatorial jet and the surrounding retrograde jet from expanding across the entire simulated region, that is, extending from the equator to the pole (Figure 2). To demonstrate this, we present four different simulations that are calculated with a set of identical control parameters, and free slip boundary conditions, each extending to a different depth (denoted by $r_{min}$, which represents the distance from the planet's center to its bottom boundary, expressed as a ratio compared to the planet's radius). The background profiles are adjusted according to the domain depth (Figure S2 in Supporting Information S1). A pair of prograde jets at midlatitudes appear slightly in the domain of $r_{min}$ = 0.84 R (Figure 2d), followed by very weak alternating jets, allowing the examination of midlatitude phenomena. The deeper domains (Figures 2a–2c), consist of no midlatitude dynamics as the equatorial region dynamics essentially extends to high latitudes and is replaced by very weak to no high-latitude jets. The dynamics in the equatorial region is affected by the position of the tangent cylinder (Heimpel & Aurnou, 2007). In deep domains, the tangent cylinder is closer to the rotation axis, leaving no place for midlatitude dynamics to evolve. Hence, to allow midlatitudinal dynamics, we must examine relatively shallow domains (e.g., Heimpel et al., 2005), or separate the outer shell dynamics from the interior via, for example, a varying electrical conductivity profile (e.g., Dietrich & Jones, 2018; Gastine et al., 2014; Heimpel & Gómez Pérez, 2011; Wicht et al., 2019; Wulff et al., 2022). Motivated by the Jupiter and Saturn gravity measurements, implying that the atmospheres are relatively shallow, we focus on the shallowest domain for the chosen control parameters (i.e., Figure 2d). Reducing the domain size causes the experiment to be numerically unstable and requires enhancement of the Ekman number (to allow numerical stability), which leads to the disappearance of the midlatitude jets (see Figure S5 in Supporting Information S1). The chosen domain depth ($r_{min}$ = 0.84R) is equivalent to the approximated radial depth of the dynamical region of Saturn (Galanti et al., 2019).

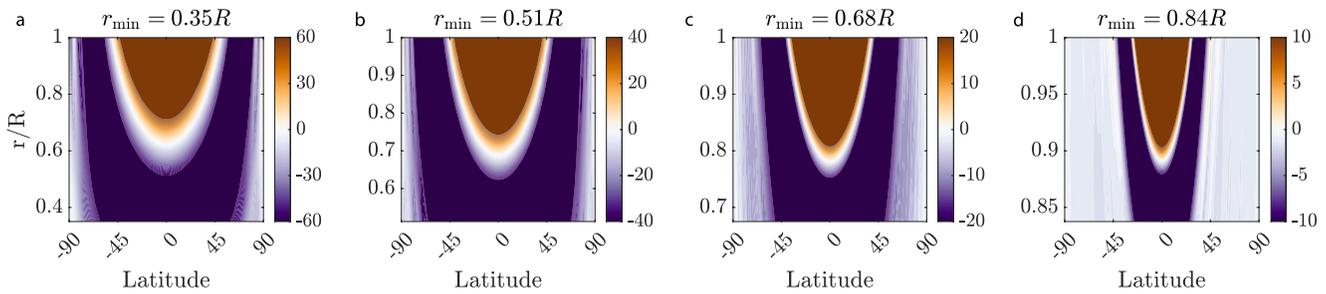

**Figure 2.** Zonally and temporally averaged zonal wind [m s$^{-1}$] of 4 simulations, each extending to a different depth ($r_{min}$). All the simulations are calculated with identical control parameters and free-slip boundaries. Note that the colors are saturated such that off equatorial jets are apparent. Zonal wind maximal and minimal values are (a) 250 and −103, (b) 193 and −91, (c) 158 and −64, (d) 90 and −38 [m s$^{-1}$].

The boundaries of the dynamical region in gas giants are influenced by the adjacent layers. Beneath the zonal flows, a bottom drag probably acts to dissipate the flow field in the form of a stably stratified layer, Ohmic dissipation in a conducting layer, wave dissipation, or a combination of these or other mechanisms (e.g., Christensen et al., 2020; Gastine & Wicht, 2021; Ingersoll et al., 2021; Liu et al., 2008). Above the zonal flows, dissipation can also occur through processes like an Ekman layer through vanishing stress or upward vertical wave propagation (e.g., Busse, 1983; Gierasch et al., 1986). These can be simulated, in a simplified manner, by applying different boundary conditions, which are known to affect the appearance and maintenance of midlatitude jets (MLJ) (Aurnou & Heimpel, 2004). Different studies conducted experiments with a variety of control parameters and boundary conditions (e.g., Aurnou & Heimpel, 2004; Jones & Kuzanyan, 2009; Wulff et al., 2022), and as can be expected, no-slip boundaries (**u** = 0 at the boundary) usually lessen the strength of zonal jets. Here, we focus on the emergence of the MLJ with different boundary conditions, all with bottom domain set to $r_{min}$ = 0.84 R, as on Saturn, and identical control parameters and reference state (See Supporting Information S1). As the no-slip upper boundary suppresses the equatorial jet, the MLJ become more pronounced. The boundary conditions also affect the vertical nature of the MLJ, as well as the width and strength of the subequatorial retrograde jets (Figure S3 in Supporting Information S1). In the no-slip top boundary scenario, two alternating, rather barotropic jets appear in both hemispheres, poleward to the subequatorial jets (Figure S3c in Supporting Information S1). Higher latitudes jets appear more prominently in the double no-slip boundaries (Figure S3d in Supporting Information S1, Figure 3), to reach a total of about 10 midlatitude alternating jets in each hemisphere (between latitudes 30°–60° S/N). The MLJ width is approximately ∼3° latitude, roughly similar to the jets in the Jovian midlatitudes (Tollefson et al., 2017).

Ferrel-like cell dynamics suggest a direct connection between the MJL and the convergence of eddy momentum fluxes, as well as the meridional streamfunction (for the mathematical definition, see Supporting Information S1) through the influence of the Coriolis force (see Section 2.2 and Equation 8). The direction of the jet should change in response to variations in eddy momentum convergence, and correspondingly, the orientation of the Ferrel cell mean circulation should also flip. By examining the meridional streamfunction under various boundary conditions (Figure S3 in Supporting Information S1, bottom row), we observe a consistent alignment with the MLJ, extending in the direction of the planetary rotation axis. Each jet is associated with a meridional streamfunction that changes its direction as the jets alternate. This alignment is particularly noticeable when employing double no-slip boundaries (Figure S3h in Supporting Information S1), as no-slip boundary essentially takes the role of eddy dissipation, or friction, near the boundaries. Further elaboration on this alignment can be found in Section 3.3. This alignment pattern is consistent with the observed deep jets and meridional circulation on Jupiter (Duer et al., 2021; Fletcher et al., 2021).

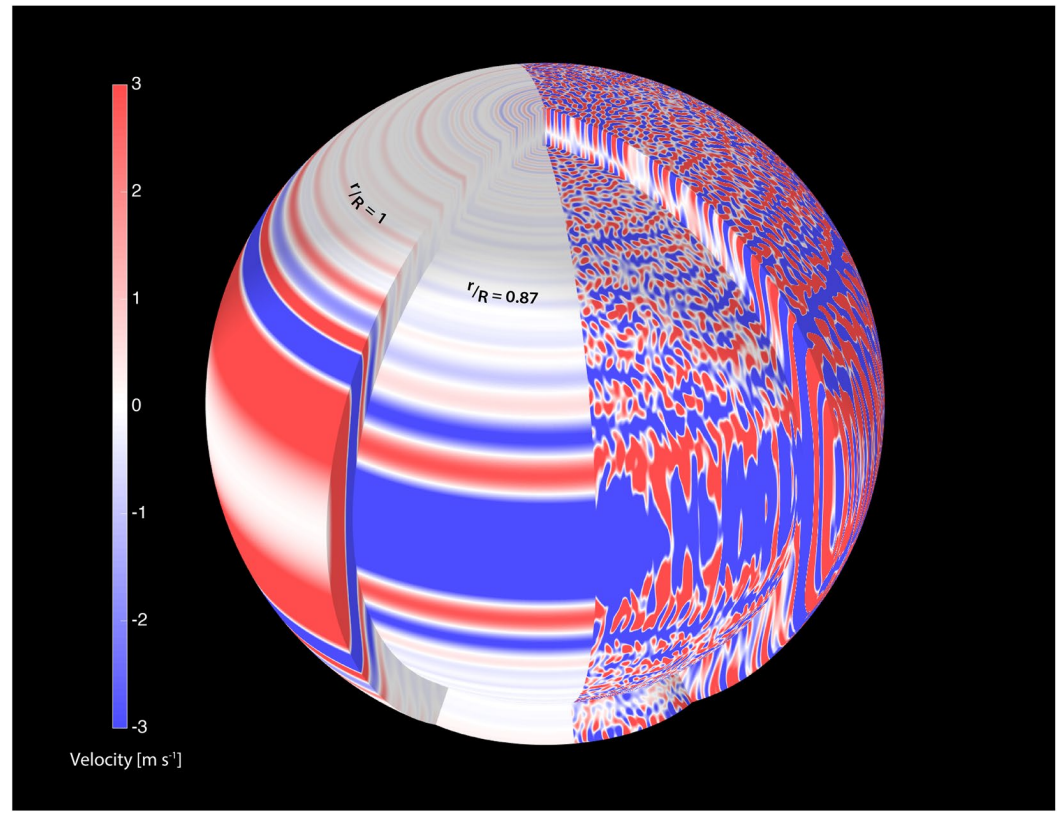

**Figure 3.** Zonal and time averaged zonal wind field (left) and a zonal wind field snapshot (right) shown at two shell depths: near the upper boundary ($r/R = 0.997 \approx 1$) and at $r/R = 0.87$ (boundary conditions are no-slip at both boundaries). Slices reveal the cylindrical orientation of the zonal wind field, existing throughout the entire domain.

To exemplify the turbulent nature of the simulation, we examine its steady state zonally-averaged wind field, together with a simulation snapshot. Figure 3 shows the former on the left and the latter on the right sides of the same multi-layered sphere. The wind field is shown in two layers, adjacent to the outer boundary ($r/R = 0.997 \approx 1$) and close to the inner boundary ($r = 0.87$ R), where the equatorial prograde flow vanishes. In the slices, the cylindrical orientation of the flow at all latitudes is visible, due to the fast rotation used in this simulation (Busse, 1976; Busse & Hood, 1982), and consistent with recent observations indicating the cylindrical nature of the zonal flows on Jupiter (Kaspi et al., 2023). The turbulence is distinguishable in the snapshot, but it also encompasses the arrangement of the zonal jets. The steady state zonally-averaged zonal wind is also presented in a line plot such that the saturated parts are apparent (Figure S5 in Supporting Information S1).

Eddy-driven jets are a well-known and studied concept, mostly in the context of the terrestrial atmosphere (e.g., Chemke & Kaspi, 2015; Dritschel & McIntyre, 2008; Flierl, 1987; Lee & Kim, 2003). However, for simulations of deep atmospheres of gas giants, the generation mechanism of zonal jets in the midlatitudes is not well-established and the other components of the velocity field (and the meridional circulation) are infrequently addressed. We examine the origin of the MLJ in this simulation (Figure 3), and whether they are eddy-driven, as are the jets in the midlatitudes of Earth. For that, we examine the first two terms of Equation 7, which are the meridional and radial eddy momentum convergence terms, along with the location of the zonal jets. The properties of the MLJ simulated here are similar, but clearly not identical compared to jets measured at the Jovian cloud layer (Figure 1a). The jets' strength is generally weaker, but could represent the velocity in the deeper regions of the atmosphere (Figure 1c). The jets' width is slightly narrower, and a clear trend of decay exists from the tangent cylinder polewards, which is stronger than the equivalent trend in the observations. This latitudinal decay is proportional to the depth of the column where eddies could evolve, which we will elaborate on next.

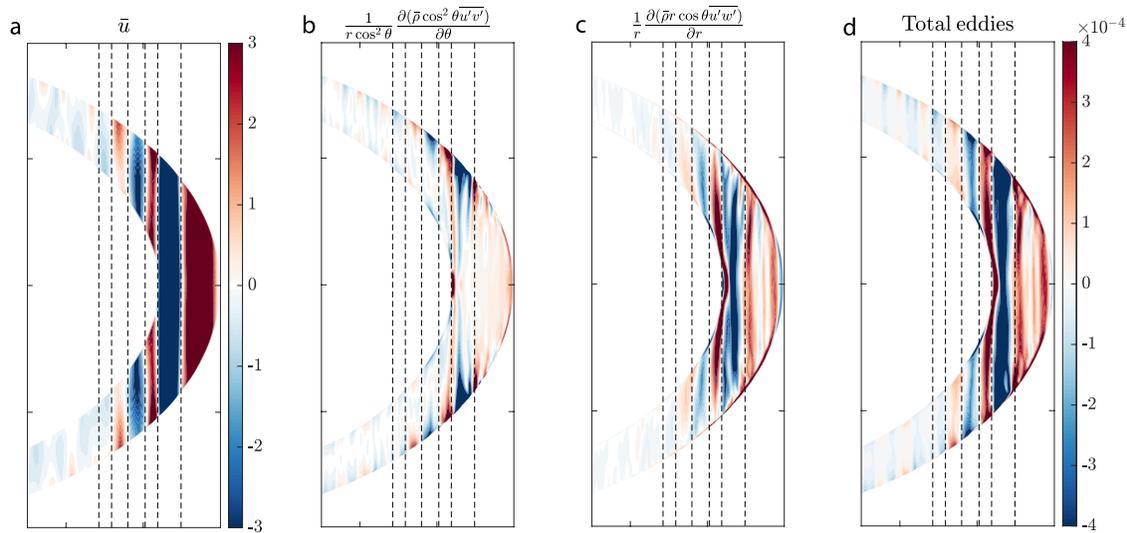

**Figure 4.** The leading order eddy terms in the simulation associated with the alternating midlatitude zonal jets. (a) The zonally averaged zonal wind [m s$^{-1}$]. (b) Meridional eddy momentum flux convergence, (c) Radial eddy momentum flux convergence, and (d) their summation [kg m$^{-2}$ s$^{-2}$]. Colorbar is shared for panels (b–d). The dashed black lines are set where the zonal velocity (panel a) is zero close to the outer boundary. The panels are showing −60° to 60° degrees latitude at the inner boundary.

The jets (Figure 4a), eddy fluxes (Figures 4b and 4c), and the meridional circulation (streamfunction, Figure S3h in Supporting Information S1) are strongly correlated with each other and exemplify the momentum relations discussed in Section 2.2, and specifically in Equation 7. As apparent in Figure 4, the radial eddy flux term is just as significant for the presence of the jets as the meridional eddy term. It appears that while the meridional eddy fluxes are dominant at the top and slightly at the bottom branches of the cells, the radial eddy momentum fluxes maintain the barotropic nature of the jet in the middle atmosphere and in the upward and downward branches of the cell. The meridional eddy fluxes (Figure 4b) reach down to a depth of approximately 2,000 km (∼1/5 of the domain), much deeper than energy constraints estimations for the real Jupiter (Liu & Schneider, 2010). However, this estimation did not consider vertical density variations, and no measurements has been able to confirm or disprove it. Nonetheless, it is clear that the meridional eddy fluxes are a main contributor for the location and pattern of the alternating jet streams in the midlatitudes. The strength of the MLJ is proportional to the column's depth (and hence, to latitude), suggesting that more eddies evolve in deeper columns, giving momentum to generate stronger jets. This is visible via the radial component of the eddy fluxes, inside the tangent cylinder (Figure 4c). The radial eddy fluxes also have a significant contribution outside the tangent cylinder, where they are significant throughout and are responsible for the tilted convection columns, aligned with the axis of rotation (Kaspi et al., 2009). Alternatively, we examine the eddy momentum fluxes in the directions perpendicular and parallel to the axis of rotation (Figure S6 in Supporting Information S1). Both at midlatitudes and in the equatorial region, the eddy fluxes in the perpendicular direction are dominant, but parallel fluxes are also significant in some regions, equivalent to the contribution of the meridional and radial fluxes in Figure 4 (see Supporting Information S1).

The total eddy momentum term (Figure 4d), the summation of the two eddy terms, gives a pattern that is equivalent in position (dashed lines in Figure 4) and relative strength (colors in Figure 4) to the zonal jets, especially in the midlatitudes. This emphasizes the importance of the two independent terms, unlike the terrestrial atmosphere, which does not contain a significant radial (or vertical, in cartesian coordinates) eddy momentum flux term. Eddies in a gas giant with a varying density and a deep enough atmosphere will hence converge in both the meridional and vertical directions, which might lead to the barotropic nature of the MLJ. These results complement other studies that explored the relationship between eddy fluxes and zonal jets in gas giant atmospheres (e.g., Christensen et al., 2020; Young et al., 2019a), but here, we reveal the importance of the radial fluxes in deep atmospheres, which are generally overlooked. In a broader context, this highlights the cylindrical nature of the flow, where the natural convergence of momentum occurs in a direction perpendicular to the planet's spin axis. This concept is essentially an extension or generalization of the standard terrestrial shallow situation, where the vertical direction is degenerate.

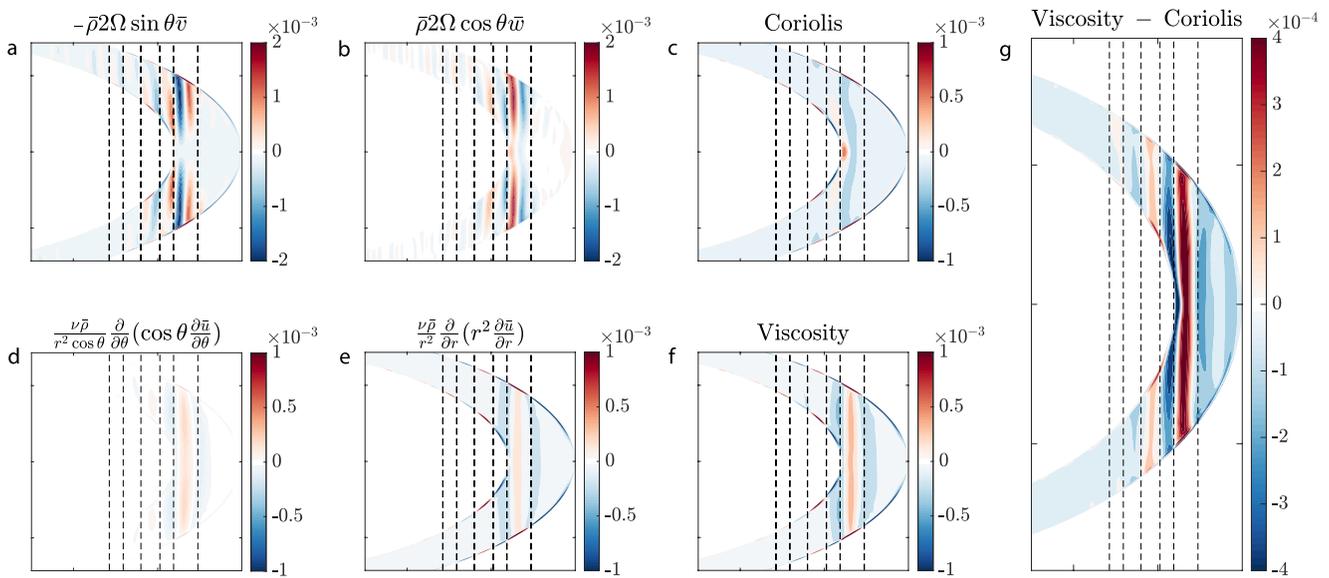

**Figure 5.** Coriolis and viscosity terms [kg m$^{-2}$ s$^{-2}$] in the zonally-averaged zonal momentum equation for the same simulation as in Figure 4. (a) and (b) The two Coriolis terms, which are close to balancing each other away from the boundaries. (c) The total Coriolis force from (a) + (b). (d) and (e) The two viscosity terms. (f) The total viscosity received from (e) + (d). The Coriolis force (c) and viscosity (f) balance each other almost entirely at the boundaries. (g) The residual to be obtained by removing (c) from (f), which is balanced by the total eddies term shown in Figure 4d (the *x*-axis and *y*-axis ranges and the colorbar are identical between panel (g) and Figure 4d). The dashed black lines are set where the zonal velocity is zero close to the outer boundary, the same as in Figure 4. (a–f) The panels are showing −60° to 60° degrees latitude (inner boundary). Note the different colorbars between panels.

### 3.2. Ferrel-Like Zonal Momentum Balance

Next, we examine all the terms of the momentum equation in order to identify the full leading order balance responsible for the maintenance of the MLJ. Figure 5 shows the additional four terms of Equation 7, that are not shown in Figure 4: two Coriolis terms and two viscosity terms. The mean momentum fluxes are small, due to the meridional and radial geostrophic velocities being small, and hence insignificant in the midlatitudes regions. Therefore, the kinetic energy is transmitted from the eddies to the mean flow. The leading order balance of Equation 7 are the two Coriolis terms, nearly balancing each other away from the boundaries (Figures 5a and 5b). This balance is the zonally averaged geostrophic balance, which is the leading order balance of the zonal momentum equation (Equation 1) for small Rossby and Ekman numbers (Kaspi et al., 2009). It also satisfies the zonally-averaged continuity equation (Equation 3). Near the boundaries, the meridional Coriolis term (Figure 5a) is balanced by the radial viscosity (Figure 5e), due to the choice of no-slip, which will necessarily cause the vertical shear of the zonal velocity to be large close to the boundaries. This balance is somewhat equivalent to the terrestrial Ferrel cell bottom boundary balance (Equation 9), where solid-surface drag is replaced here by the radial numerical viscosity. The additional (meridional) viscosity term is weaker, but still significant away from the boundaries (Figure 5d). Summing the two Coriolis terms (Figure 5c) and two viscosity terms (Figure 5f) gives a term (Figure 5g) that is balancing almost entirely the total eddy term (Figure 4d), indicating that Equation 7 holds. Unlike the terrestrial Ferrel cells, the radial eddy term plays a key role in sustaining the balance. Yet, the upper branch balance (Equation 8) is manifested in the upper region of the simulated atmosphere as well, where the Coriolis force is balanced by the meridional eddy momentum flux (away from the boundary layer). A similar mechanical balance can be found in Gastine and Wicht (2021), where the Reynolds stresses (eddy momentum flux terms) maintain the alternating zonal jets. However, the different boundary conditions and background profiles result in a different balance in the lower part of the domain. An interesting aspect highlighted in their model is the role of the Lorentz force at the lower boundary of the conductive region, above the steady stratified layer, which can effectively substitute for a bottom drag. Note that Equation 7 holds not only in midlatitudes but also in the equatorial region, manifesting the known, leading order, balance of Taylor convection columns, generated by the Reynolds stresses in the axis of rotation plane (Busse, 2002). These Taylor columns are aligned in cylinders outside the tangent cylinder, and hence are dominated by the eddies that are perpendicular to the axis of rotation (see Figure S6 in Supporting Information S1).

### 3.3. Meridional Circulation Cells

Various Earth-based and spacecraft measurements have suggested the existence of meridional circulation in Jupiter's midlatitudes (Fletcher et al., 2020). However, only recently the presence of the meridional circulation cells have been established by the Juno MWR measurements (Duer et al., 2021), driving us to study them via numerical simulations. Different boundary conditions have a direct impact on the number and structure of the meridional circulation at midlatitudes. Yet, all combinations of boundary conditions allow meridional cells that accompany the zonal jets (Figure S3 in Supporting Information S1, bottom row). Interestingly, any combination with free-slip boundary conditions lead to only one cell surrounding each jet, but the double no-slip boundary conditions reveal a stacked circulation pattern (two oppositely stacked circulations cells) aligned with the six lowest-latitude (adjacent to the subequatorial retrograde jet) MLJ (Figure S3h in Supporting Information S1).

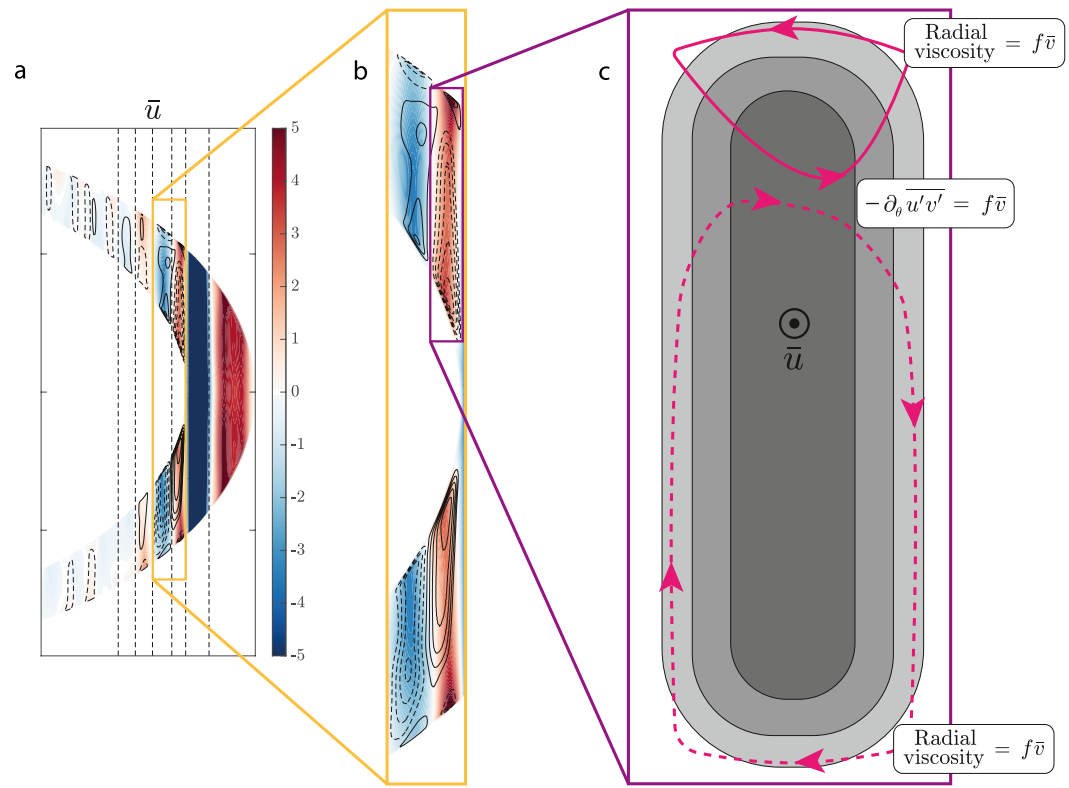

**Figure 6.** (a) The zonally averaged zonal wind (colors, [m s$^{-1}$]) and the meridional streamfunction (lines, [s$^{-1}$]). (b) A magnification of four jets and the circulation surrounding them, revealing a stacked circulation pattern. (c) An illustration of the stacked cells surrounding an eastward jet in the northern hemisphere along with the force balance at the meridional branches of the circulations (Equations 10). Solid (dashed) lines are anti-clockwise circulation and dashed (solid) lines are clockwise circulation in the northern (southern) hemisphere.

### 3.3.1. The Stacked Cells Hypothesis

The morphology of stacked circulation cells agrees with some observations of the Jovian troposphere. Temperature measurements (Fletcher et al., 2016), and tracer anomalies (de Pater et al., 2001; Gierasch et al., 1986) imply that the circulation above the 1 bar pressure level is reversed compare to the circulation identified below (Duer et al., 2021; Fletcher et al., 2021; Ingersoll et al., 2000; Showman & de Pater, 2005). Both circulations are driven by the meridional eddy fluxes measured at the cloud level, but what mechanism can close the balance at the boundaries is still a mystery, as discussed above in Section 3.1. In the morphology shown here (Figures 6a and 6b), both boundaries (top and bottom) are dominated by a balance between the Coriolis force and the radial viscosity term (Figures 5 and 6c), such that

$$\frac{\nu}{r^2}\frac{\partial}{\partial r}\left(r^2 \frac{\partial \bar{u}}{\partial r}\right) = f\bar{v}. \tag{10}$$

The radial viscosity term is large due to the imposed boundary conditions. It is unlikely that a gas giant would hold such a balance in either boundary, but it successfully demonstrates that a friction-like force would lead to a circulation adequate to the stacked circulation pattern, fitting the observations from the Jovian atmosphere. Removing either strict boundary condition leads to a more terrestrial-like circulation result, where one (top-to-bottom) cell accompanies each jet (Figure S3 in Supporting Information S1). The vertical shared branch between the stacked cells is the known Ferrel cell leading order balance, equivalent to the upper branch balance of the terrestrial Ferrel cell, where the Coriolis force is balanced by the meridional eddy momentum flux convergence (Equation 8). This balance has been shown to also dominate in the Jovian cloud-level layer (Duer et al., 2021).

## 4. Conclusions

As new spacecraft measurements are gained, the atmospheric circulation of Jupiter and Saturn is being revealed, necessitating the need for extensive analysis and studies to provide robust theoretical and numerical backing for these revelations. This study focuses on the meridional circulation pattern and deep behavior of the atmospheric jet-streams on Jupiter, which the unprecedented measurements of the Juno spacecraft have further revealed (e.g., Bolton et al., 2021). Juno measurements emphasize that the Jovian atmosphere can be divided into three regions: the equatorial eastward flows, containing a superrotation, and surrounding retrograde jets ($\sim 20°$ S $- 20°$ N); the midlatitudes alternating jet streams and the meridional circulation cells ($\sim 60°$ S $- 20°$ S and $\sim 20°$ N $- 60°$ N); and the turbulent polar region and polar cyclones ($\sim 90°$ S $- 60°$ S and $\sim 60°$ N $- 90°$ N) (Figure 7). The simulations presented here capture the first two regions, and the analysis focuses on the eddy-driven, midlatitudinal, alternating jets and the Ferrel circulation cells surrounding them.

The MLJ are driven by turbulence in the atmosphere, but unlike the MLJ in the shallow terrestrial atmosphere, where the vertical direction is degenerate, their barotropic nature is dictated by both meridional and radial fluxes, each dominant at a different depth (Figure 4). This emphasizes the cylindrical orientation of the flow, where momentum predominantly converges in a direction perpendicular to the axis of rotation. Rewriting the meridional and vertical fluxes as fluxes parallel and perpendicular to the axis of rotation shows the dominance of the perpendicular fluxes in most regions in the interior (Figure S6 in Supporting Information S1). This implies that the standard dominance of the meridional fluxes in the terrestrial atmosphere is a particular case where the jet aspect ratio is large (as on Earth). The Coriolis force and numerical viscosity are also accounted for when considering the full zonal momentum equation, resulting in circulation that is somewhat equivalent to the Ferrel cell circulation, with the relevant adjustments to a gas giant (Figure 5, Equation 7). The eddy-driven midlatitudes dynamics are a direct result of the chosen control parameters. They do not emerge in deep domains where the equatorial dynamics occupy most latitudes, extending toward the poles (Figure 2). The numerical viscosity also must be kept at relatively small values. Otherwise, off-equatorial eddy-related phenomena vanish (Figure S5 in Supporting Information S1).

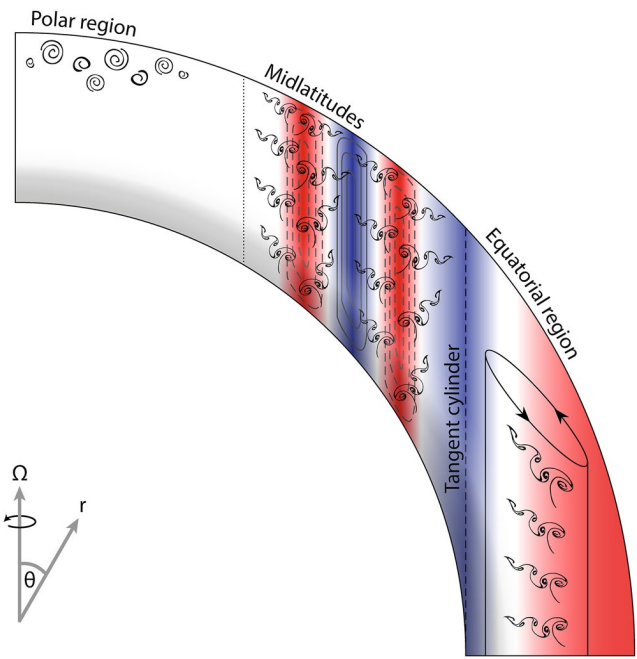

**Figure 7.** Illustration of Jupiter's three distinct atmospheric regions: the equatorial eastward flows and surrounding retrograde jets; the midlatitudes, with the eddy-driven, alternating jet-streams and meridional circulation cells (Duer et al., 2021; Salyk et al., 2006); and the turbulent polar region (Adriani et al., 2018; Gavriel & Kaspi, 2021).

This study does not aim to explain the dissipating mechanism of the zonal jets, but it replaces it with changing set of boundary conditions, which imitate the potential effects on the zonal jets (Figure S3 in Supporting Information S1). Nonetheless, it captures well multiple features that are equivalent to the Jovian atmosphere and provides dynamical support for them, like the non-forced eddy-driven jets and the stacked meridional circulation pattern (Figure 6). These results likely apply to the Saturnian atmosphere as well, as the MLJ there are also eddy-driven at the cloud level (Figure 1), and the atmosphere is deep enough such that radial eddy fluxes are to be considered, but not too deep such that turbulent midlatitudes are possible.

## Conflict of Interest

The authors declare no conflicts of interest relevant to this study.

## Data Availability Statement

The data presented in the introduction part of this study were published by Tollefson et al. (2017), Garcia-Melendo et al. (2011), Salyk et al. (2006), Del Genio and Barbara (2012), Kaspi et al. (2018), Galanti et al. (2019), and Galanti and Kaspi (2021) as cited in the text. The numerical results were obtained by solving the hydrodynamic equations using the Rayleigh code (Featherstone et al., 2022) with the reference state and control parameters as detailed in Supporting Information S1.

**Acknowledgments**

We thank the reviewers for their important feedback and constructive comments on the earlier version of this manuscript. We acknowledge valuable assistance with implementing the Rayleigh model from Nick Featherstone and the graphic tool development contributed by Oded Aharonson. We acknowledge the support of the Israeli Space Agency, the Israeli Science Foundation (Grant 3731/21), the Helen Kimmel Center for Planetary Science at the Weizmann Institute of Science, and the Juno mission.